\documentclass[numsec,webpdf,modern,large]{oup-authoring-template}%

\graphicspath{{Fig/}}

\usepackage{xspace}
\usepackage[dvipsnames]{xcolor}
\usepackage{tikz}
\usepackage{pgf-pie}
\usepackage{balance}
\usepackage{placeins}

\theoremstyle{thmstyleone}%
\theoremstyle{thmstyletwo}%
\newtheorem{example}{Example}%
\theoremstyle{thmstylethree}%

\newcommand\YAMLcolonstyle{\color{red}\mdseries}
\newcommand\YAMLkeystyle{\color{black}\bfseries}
\newcommand\YAMLvaluestyle{\color{blue}\mdseries}

\makeatletter

\newcommand\language@yaml{yaml}

\expandafter\expandafter\expandafter\lstdefinelanguage
\expandafter{\language@yaml}
{
  keywords={true,false,null,y,n},
  keywordstyle=\color{darkgray}\bfseries,
  basicstyle=\YAMLkeystyle\scriptsize,                                 
  sensitive=false,
  comment=[l]{\#},
  morecomment=[s]{/*}{*/},
  commentstyle=\color{purple}\ttfamily,
  stringstyle=\YAMLvaluestyle\ttfamily,
  moredelim=[l][\color{orange}]{\&},
  moredelim=[l][\color{magenta}]{*},
  moredelim=**[il][\YAMLcolonstyle{:}\YAMLvaluestyle]{:},   
  morestring=[b]',
  morestring=[b]",
  literate =    {---}{{\ProcessThreeDashes}}3
                {>}{{\textcolor{red}\textgreater}}1     
                {|}{{\textcolor{red}\textbar}}1 
                {\ -\ }{{\mdseries\ -\ }}3,
}

\lst@AddToHook{EveryLine}{\ifx\lst@language\language@yaml\YAMLkeystyle\fi}
\makeatother

\newcommand{\schemalink}{SchemaLink\xspace}

\begin{document}

\journaltitle{Journal Title Here}
\DOI{DOI HERE}
\copyrightyear{2022}
\pubyear{2019}
\access{Advance Access Publication Date: Day Month Year}
\appnotes{Paper}

\firstpage{1}


\title[\schemalink: An Intelligent Web 
Editor for LinkML Schema Curation]{\schemalink: An Intelligent Web 
Editor for LinkML Schema Curation}

\author[1]{Emanuele Cavalleri\ORCID{0000-0003-1973-5712}}
\author[1]{Paolo Perlasca\ORCID{0000-0001-6674-2822}}
\author[2]{J. Harry Caufield\ORCID{0000-0001-5705-7831}}
\author[2]{Justin Reese\ORCID{0000-0002-2170-2250}}
\author[2]{Christopher J. Mungall\ORCID{0000-0002-6601-2165}}
\author[1,2,$\ast$]{Marco Mesiti\ORCID{0000-0001-5701-0080}}

\authormark{E. Cavalleri et al.}

\address[1]{\orgdiv{Department of Computer Science}, \orgname{University of Milano}, \orgaddress{\street{Via Celoria 18, Milano}, \postcode{20133}, 
\country{Italy}}}
\address[2]{\orgdiv{Biosystems Data Science}, \orgname{Lawrence Berkeley National Lab}, \orgaddress{\street{Calvin Rd, Berkeley}, \postcode{94705}, \state{CA}, \country{USA}}}

\corresp[$\ast$]{Corresponding author. \href{email:marco.mesiti@unimi.it}{marco.mesiti@unimi.it}}

\received{Date}{0}{Year}
\revised{Date}{0}{Year}
\accepted{Date}{0}{Year}


\abstract{
\textbf{Motivation:} LinkML is a suitable language for the representation of the structural and content constraints of different kinds of biomedical data. Even if it is a quite recent proposal, it has been applied in several biomedical contexts.
Developing and maintaining LinkML schemas presents several challenges, particularly for novice curators. Non-expert bio-curators may struggle with LinkML syntax and best practices, requiring significant time and effort to develop well-structured schemas.\\
\textbf{Results:} In this paper we propose SchemaLink, a web-based environment for the graphical construction and enhancement of LinkML schemas that address the following requirements: $(i)$ introduce a graphical language for the specification of LinkML schemas, $(ii)$ make uniform the specification of schemas in similar contexts,~$(iii)$~simplify the design and curation processes by exploiting a RAG-based approach to assist curators in creating new schemas from scratch and editing already developed ones. Several experimental analyses 
show the quality of the produced LinkML schemas through the AI-based editing facilities.\\
\textbf{Availability and Implementation:} SchemaLink is available online at: {\tt https://SchemaLink.biodata.di.unimi.it}. SchemaLink code and testing data are available as open-source on GitHub at: {\tt https://github.com/AnacletoLAB/}$\{${\tt schemalink-webapp},{\tt schemalink-api}$\}$.\\
}


\maketitle

\section{Introduction}
LinkML \citep{linkml} is a flexible modeling language for creating YAML-based schemas 
across various biomedical domains. 
It allows the precise specification of biomedical entities with their relationships and distinctive properties. 
Unlike OBO ontologies~\citep{obo} and biomedical controlled vocabularies (e.g. KEGG~\citep{kegg}, Rfam~\citep{rfam}) that characterize entire domains of interest, LinkML schemas focus on specific aspects that are of interest for conducting a given analysis.  
Classes and relationships can be grounded in ontologies to represent their semantics and leverage their identification schemes.
Descriptions and instances of schema elements can be included to improve the characterization of key entities and document their roles.
Although it is a relatively recent framework, LinkML has already been applied in several biomedical contexts, including the identification and extraction of biomedical entities and relationships~\citep{CavalleriGPMCRM24}, cancer data harmonization~\citep{cancerlinkml}, environmental genomics~\citep{omicslinkml}, and knowledge graph integration~\citep{Unni2022}.

Developing and maintaining LinkML schemas, however, presents several challenges, particularly for novice curators. Non-expert bio-curators may struggle with LinkML syntax and best practices, requiring significant time and effort to develop well-structured schemas. Moreover, the following issues impact LinkML schema curation:
($i$)  
The richness and flexibility of LinkML  
offer diverse tree- or graph-like
schema structural forms,
which can lead to inconsistencies, making schemas difficult to compare and integrate across similar application domains.
($ii$) Understanding and validating existing LinkML schemas typically involves manually inspecting hundreds of lines of code, which is both time-consuming and error-prone. Updating and versioning schemas may introduce inconsistencies.
($iii$) Classes and properties imported from biomedical ontologies may be subject to semantic ambiguity. Misalignment between ontology terms and LinkML classes can introduce both syntactic and semantic errors, complicating schema interoperability. 

\begin{figure*}[t]
    \centering    \hspace*{2cm}\includegraphics[width=.7\linewidth]{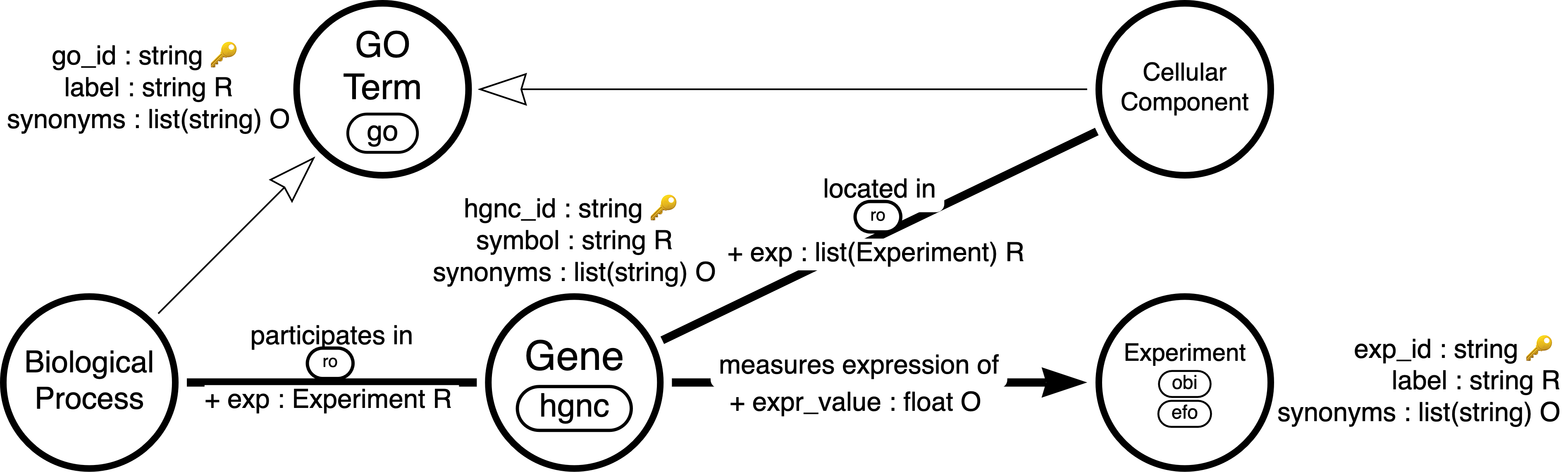}
    \caption{Graphical representation of a schema.}
    \label{fig:schema}
\end{figure*}

In this paper, we introduce SchemaLink, a web-based environment for the graphical construction and enhancement of LinkML schemas that address the previously outlined challenges.  
SchemaLink extends the {\tt arrows.app} \citep{arrows} graphical environment proposed by Neo4j to develop property graphs. While {\tt arrows.app} is intended to draw graph nodes and relationships at the instance-level, SchemaLink has been conceived to work at the schema-level, allowing schema curators to define classes and specify associative and inheritance relationships among them. 
Beyond class and relationship modeling, SchemaLink enables the definition of various schema constraints, including attribute types, primary keys, mandatory and optional properties, cardinality constraints on attributes and relationships, default values, controlled vocabularies, descriptions, and examples taken from ontologies.
All these characteristics can be  
exploited via simple graphical artifacts, avoiding the use of LinkML syntax. 
Validation mechanisms are included in the platform to check the adherence of the schemas to LinkML syntax. Generated graphical schemas can then be serialized according to different schema structural forms relying on the user's needs. 

An intelligent component has been integrated in SchemaLink that exploits a vector database, fed with LinkML schemas developed 
in several Monarch projects~\citep{monarch}, and general-purpose large language models (LLMs) combined with a 
Retrieval-Augmented Generation (RAG) approach~\citep{rag} to assist curators in creating new schemas from scratch and enhancing already developed ones. For refining a schema $S$, different kinds of schema modification prompts have been formulated (like adding a class, including new relations between two classes, explaining the role of a class/relation in the schema) that can be issued to a LLM along with schema samples similar to $S$ (extracted from the vector database) to suggest a new version that includes the proposed modification. This is an innovative contribution of SchemaLink that can be exploited both by novice and expert LinkML curators. 

Several tools, proposed by different research communities, provide a graphical abstraction of the schema adopted in a given domain. Besides the general purpose modeling systems (e.g. Entity-Relationship diagrams, UML class diagrams, and ontology editor frameworks such as Prot\'eg\'e~\citep{protege}), specific ones have been introduced to describe schemas. Approaches can be classified into three main paradigms: Labeled-property-graph (LPG) diagrams~\citep{LPG}, Object-role modeling (ORM) diagrams~\citep{Halpin2007}, Ontology-based diagrams (e.g. OntoPad~\citep{Arndt2021AVS}, SHAPEness~\citep{Paciello23}). These models provide different expressive powers for capturing schema constraints. However, to the best of our knowledge, none of them incorporates artificial intelligence features to assist end-users in designing schemas, nor they provide support for LinkML structural forms.

A key requirement that should be guaranteed by an intelligent application like the one we propose is that the proposed suggestions are meaningful and lead to identifying a reasonable schema that fulfills the user's expectations. However, identifying expert curators to calibrate the system is often unfeasible in real-world settings, due to both the limited availability of qualified experts and the costs associated with their involvement. For this reason, in this paper, we envisioned the use of the LLM-as-a-Judge~\citep{llmjudge} technique for evaluating the quality of the obtained schemas. This approach allowed us to conduct extensive testing of different use cases and to limit the use of domain experts to a very limited subset of questionable cases.

\section{System and Methods} 
\schemalink relies on an abstract graph data model according to which LinkML schemas can be easily represented. The use of graphical artifacts for representing the schema components makes easy the exploration and development of LinkML schemas. Moreover, an intelligent component has been realized to further support the curators in the design and enhancement of schemas that exploits a RAG-based method.  Furthermore, 
our schemas can be easily serialized in different structural forms, according to which real-world LinkML schemas are generated.

\begin{figure}[t]
    \centering
    \includegraphics[width=0.63\linewidth]{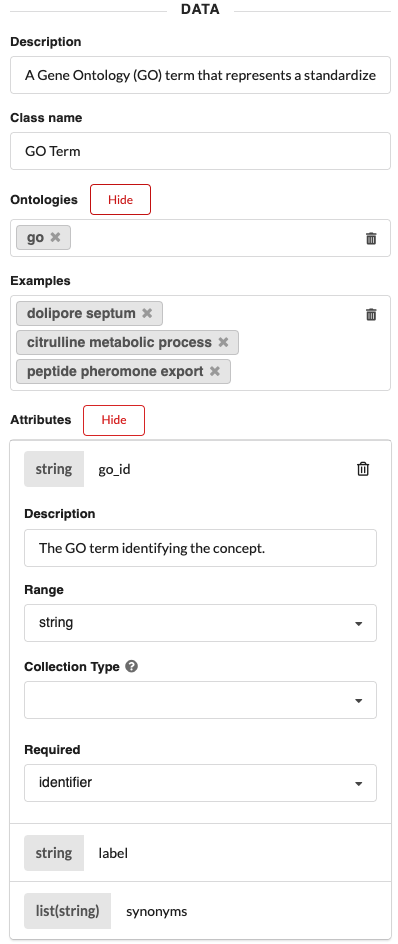}
    \caption{Graphical artifacts for the specification of classes.}
    \label{fig:classes}
\end{figure}

\subsection{A Graph Model for LinkML Schemas}\label{sec:model}
A schema $S$ is a triple $\langle \mathcal{C},\mathcal{R}, \mathcal{M}\rangle$, where 
$\mathcal{C}$ and $\mathcal{R}$ are the set of classes and relationships of the schema with the description of their attributes, and $\mathcal{M}$ is a set of metadata associated with the schema. \schemalink allows the graphical specification of a schema $S$ by means of different graphical artifacts that are described in this section. The graphical representation is internally translated into a LinkML schema because its textual representation facilitates the use of LLMs for suggesting schema modifications. 

\begin{example}
Suppose a biomedical curator wishes to integrate -omics data from heterogeneous sources structuring knowledge for downstream tasks such as differential expression and pathway analysis and functional enrichment studies. For this purpose, the curator needs to create a schema for representing genes participating in specific biological processes and their localization within cellular components (annotations can be retrieved from Gene Ontology terms~\citep{go}). Moreover, gene expression can be measured via wet lab experiments such as PCR, microarrays, and RNA-seq.
For the generation of this schema, he can exploit \schemalink and develop the diagram in Fig.~\ref{fig:schema}.
\end{example}

\par{\bf Classes.} 
A class $C\in \mathcal{C}$ is visually represented as a circle that can be added to the main canvas.
When the circle is highlighted an inspector panel is opened on the right side of the canvas (as the one in Fig.~\ref{fig:classes}). The user can specify the class name (denoted {\tt class\_name}) unique in the entire schema, a description, ontologies to annotate the class with, and a sample of class instances (the latter can be used for entity extraction through few-shot and LLMs~\citep{spires}). The backend application supports the user in the specification of OBO ontologies and in suggesting samples of OBO terms to be used as examples through the Ontology Lookup Service API~\citep{ols}. Through the inspector, the user can manage the class attributes.
For each attribute, the name should be declared along with a textual description and its type (simple types like {\tt int}, {\tt bool}, {\tt string}, {\tt date}, composite types like {\tt list}, {\tt set}, {\tt tuple} of simple types, Regex-based types for pattern validation). 
Alternatively, attributes can reference other classes. This feature of embedding classes as attributes allows the specification of inlined instances in models that support this functionality~\citep{linkml}. Attributes can also be marked as mandatory and used as identifiers.
Fig.~\ref{fig:classes} shows the inspector panel for the {\tt GO Term} class of our running example. For this class, the curator has identified the properties {\tt go\_id}, {\tt label}, and a list of {\tt synonyms}.

For the sake of readability, the canvas reports the
class name, the ontologies associated with it, and the basic properties of its attributes.  
The inspector panel reports details with the visualization options for drawing the class and its properties. 

\begin{figure}
    \centering
    \includegraphics[width=0.65\linewidth]{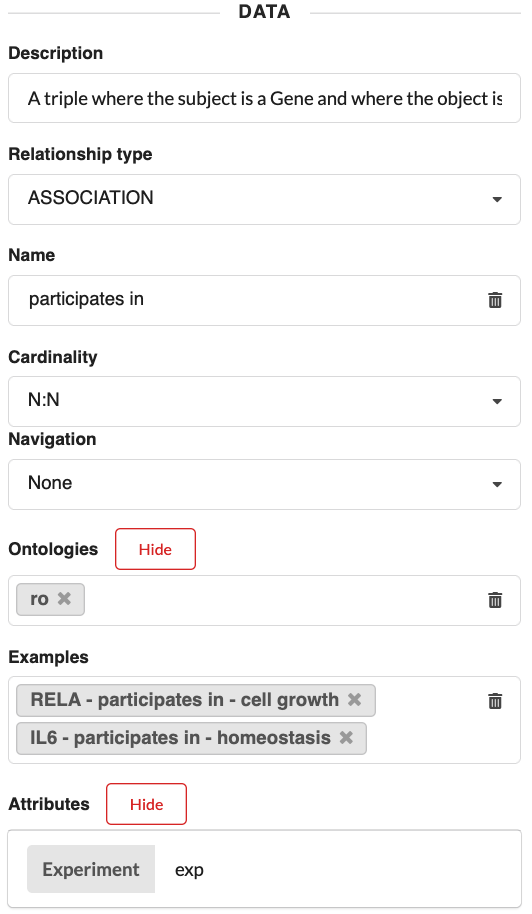}
    \caption{Graphical artifacts for the specification of associations.}
    \label{fig:associations}
\end{figure}

\par{\bf Association and Inheritance Relations.} 
A class can be linked to another one to draw a relation $R \in \mathcal{R}$ by hovering outside a node and dragging the cursor. This feature is inherited from the {\tt arrows.app} application and permits to intuitively connect classes and build schemas.
Then, the panel inspector can be shown to specify the relationship type. By default, a link between two classes is an association relationship, and the user can specify its name (denoted {\tt rel\_name}) and cardinality constraints ({\em zero-to-one}, {\em one-to-one}, {\em zero-to-many}, {\em one-to-many}, {\em many-to-many}, as well as the option of defining a custom minimum and maximum cardinality for both the subject and the object). Depending on the adopted cardinality constraint, a different kind of arrow is drawn. 
The user may also report the way in which the relationship should be navigated when serializing the schema (from left to right, from right to left, or both directions). 
Moreover, the user can annotate the relationship using ontologies that formally define the relation type and retrieve samples of relationships from the considered ontologies (backend facilities similar to the one described for classes are available here). Finally, relationships can be characterized through attributes and their specification is similar to the one for classes. In this case, attributes of type class are used to represent $n$-ary relationships (given its LPG roots, \schemalink facilitates the specification of binary relationships while offering flexibility to other relationship types).
Fig.~\ref{fig:associations} shows the panel inspector for the association {\tt participates in} between the classes {\tt Gene} and {\tt BiologicalProcess}. For this association, the curator has identified an instance sample of plausible relationships and an attribute named {\tt exp} referencing the class {\tt Experiment}. The name of the relationship (i.e. {\tt participates in}) is grounded in the corresponding Relation Ontology (RO~\citep{ro}) property.

Once an association is drawn, it can be updated into an inheritance one that is rendered as a UML-like empty arrow. When this option is chosen, the inspector panel is simplified by maintaining only the relation type. 
Once the inheritance relationship is specified, the child class inherits all the properties and associations of the parent class.
Fig.~\ref{fig:schema} shows that {\tt Bio\-logicalProcess} and {\tt CellularComponent} are subclasses of {\tt GoTerm}.

\begin{figure}[t]
    \centering
    \includegraphics[width=\linewidth]{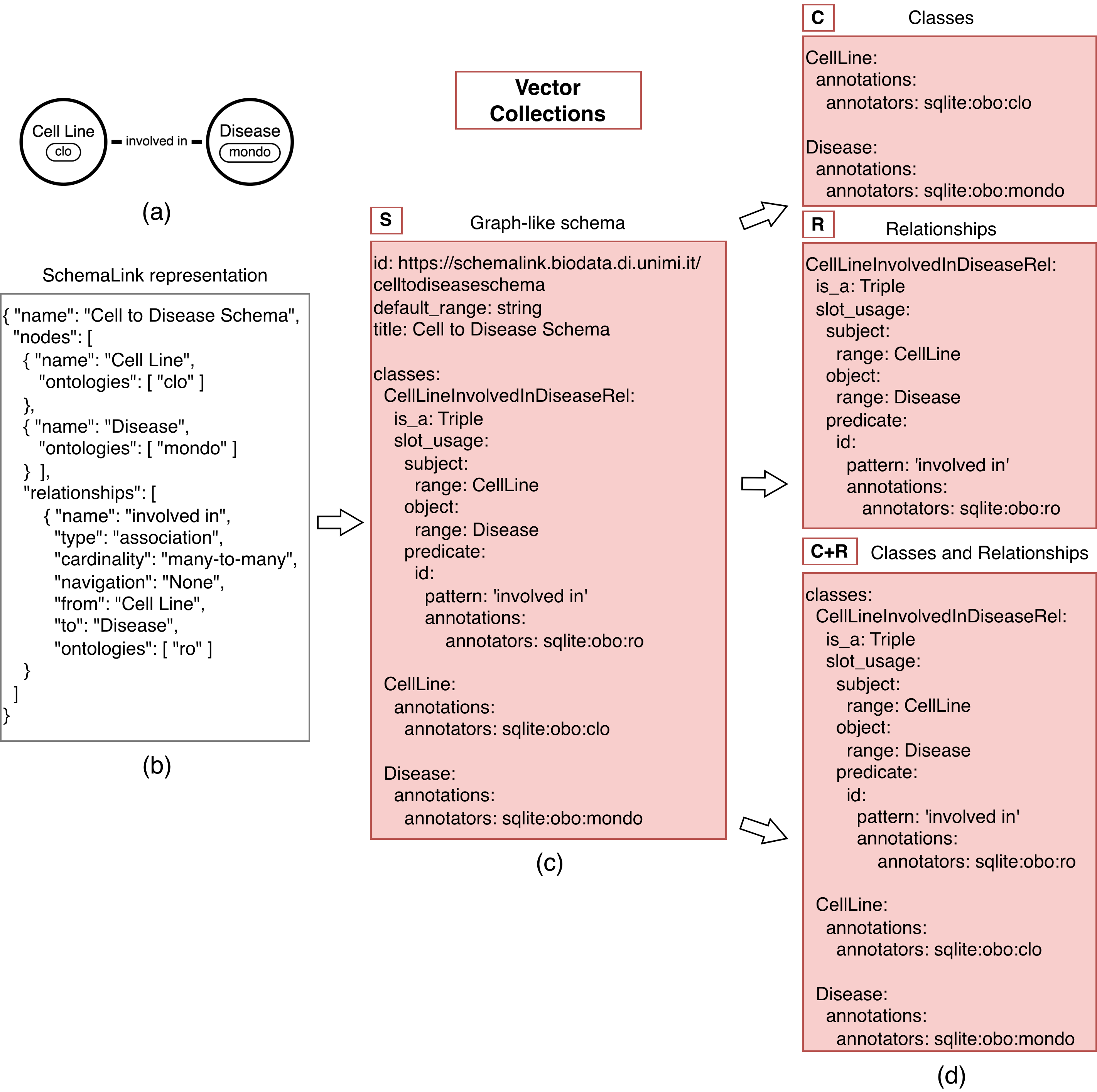}
    \caption{Internal representation of LinkML schemas and schema portions for feeding the collections of the vector database.}
    \label{fig:oovsrdflinkml}
\end{figure}

\subsection{The \schemalink Intelligent Component}\label{sec:editor}

\schemalink integrates a RAG approach to support both the semi-automatic generation of schemas from scratch and the enhancement of existing schemas through a structured set of parametric schema editing prompts. These prompts, instantiated on the selected schema items, guide a general-purpose LLM in suggesting coherent extensions, refinements, or explanations consistent with the existing schema structure.

When generating a schema from scratch, the user provides a textual description of the target domain. Based on this description, the system proposes an initial LinkML schema in the editing canvas, including candidate classes and relationships. The user may then refine the schema either manually (by interacting with the graphical artifacts) or by invoking the intelligent editing functionality to apply modification operations to the entire schema or to selected portions of it. 

In the remainder, we describe the organization of the vector database, the taxonomy of schema modification prompts, and the overall workflow of the intelligent \schemalink component.

\par{\bf Vector Database.}
The vector database is organized into the following four 
collections, 
designed to support specific schema editing operations during the RAG:
$(i)$ entire LinkML schemas (collection $(S)$);
$(ii)$ only classes extracted  (collection $(C)$);
$(iii)$ only relationships (collection $(R)$); and
$(iv)$ relationships enhanced with the specifications of the classes they connect (collection $(C\!+\!R)$). Examples of schemas are initially introduced in the $(S)$ collection, while the other collections are automatically populated. 
A Levenshtein distance~\citep{leven} filter is used to avoid the presence of near-duplicate classes and relationships within a collection (details in the Supplementary Materials).

\begin{example}
Consider the simple schema reported in Fig.~\ref{fig:oovsrdflinkml}.a that describes the cell lines involved in diseases. Fig.~\ref{fig:oovsrdflinkml}.b shows the internal representation within \schemalink, whereas  Fig.~\ref{fig:oovsrdflinkml}.c shows the corresponding graph-like form that is stored in the collection $(S)$. Finally, Fig.~\ref{fig:oovsrdflinkml}.d shows the portions of schemas that are stored in the other collections.  
\end{example}

\par{\bf Schema Editing Operations.}
\schemalink offers a set of 43 schema editing operations. These operations have been classified according to two dimensions:~$(i)$~the target of the modification, i.e. the portion of the schema to which the operation applies (a class, relationship, subgraph involving multiple classes and relationships), and $(ii)$ the type of operation performed on that target. 
Four operation types have been identified: {\tt Add}, which includes operations that introduce new classes/relationships/attributes or annotate existing ones; {\tt Fix}, which encompasses refinements aimed at improving the quality of existing schema items, including enhancements to class names, attribute types, and cardinality of relationships; {\tt Explain}, which provides human-readable explanations of the role and semantics of a selected schema portion within the overall schema; {\tt Reification}, which includes operations that promote attributes from a class to new classes linked through appropriate relationships.

Moreover, we define a repository of schema editing prompts containing pairs $\langle op, M_{op} \rangle$, where $op$ denotes an editing operation and $M_{op}$ is its associated parametric prompt template. 
At run-time, the template $M_{op}$ is instantiated according to the selected schema portion (i.e. the class, relationship, or subgraph selected by the user) and associated parameters (including the element name, attributes, reference ontologies, and instance sample).

Listing~\ref{lst:addassociationparametricprompt} reports the parametric template prompt associated with the operation \texttt{Add association} for the generation of 
a new association between two classes. Parameters (rendered in blue) are dynamically replaced with the names of the selected classes.

\begin{lstlisting}[caption={Parametric prompt
for generating a new association between two classes.},
                   basicstyle=\scriptsize\ttfamily,
                   escapeinside={(*}{*)},
                   label={lst:addassociationparametricprompt},
                   xrightmargin=-12pt]
From the LinkML schema provided below, add one or more new
semantically meaningful relationships between the classes
(*\textcolor{blue}{class\_name\textsubscript{1}}*) and (*\textcolor{blue}{class\_name\textsubscript{2}}*).
\end{lstlisting}

Finally, each $op$ is associated with a specific collection of the vector database, which determines the type of examples retrieved through the RAG.
The selected collection reflects the contextual needs of the operation (e.g. full-schema context, class-level patterns, relationship-level patterns, or both classes and relationships) and takes advantage of an empirical evaluation that we have conducted (details will be provided in the experiments).

Table~\ref{tab:schemaquery} summarizes the main modification operations supported by \schemalink for classes and relationships, along with their target, operation type, and associated vector database collection ({\tt Coll.}). Operations on subgraphs extend or combine those defined for classes and relationships.

\begin{table*}[t]
    \centering
    \begin{footnotesize}
    \begin{tabular}{|p{1.4cm}|p{1.7cm}|p{10.9cm}|
    c
    |}
    \hline
        {\bf Target} & {\bf Operation}  & {\bf Schema Modification} & {\bf 
       Coll.
        }  \\ \hline
        \multirow{12}{1.2cm}{{\bf class}}
         & \multirow{4}{1.3cm}{{\bf add}} &  add a new class semantically similar to {\tt class\_name} & {\tt S} \\
        &  & add a new class in relation with {\tt class\_name} through the relation {\tt rel\_name} & {\tt C} \\
        & & add [attributes$|$attributes description$|$parent class$|$child class] to {\tt class\_name} 
        & {\tt S} \\
        & & annotate  {\tt class\_name} with relevant [ontologies$|$samples$|$description] & {\tt C} \\ \cline{2-4}
        
        &  \multirow{5}{1.3cm}{{\bf fix}} & rename  {\tt class\_name} & {\tt R} \\
        & & enhance attributes [name$|$description] of {\tt class\_name}
        & {\tt R} \\
        & & update attributes type for {\tt class\_name}
        & {\tt C+R} \\
        & & update ontologies that annotate {\tt class\_name} & {\tt S} \\
        & & enhance [sample] for {\tt class\_name} & {\tt C} \\\cline{2-4}
        & \multirow{1}{1.3cm}{{\bf reification}} & extract attributes from {\tt class\_name} to create a new class & {\tt S} \\ \cline{2-4}
        & {\bf explain} &  explain in human-friendly terms the role of {\tt class\_name} in the schema & - \\ \hline    
        
\multirow{11}{1.2cm}{{\bf relation}} & \multirow{4}{1.3cm}{{\bf add}} &  add a new association between  {\tt class\_name}$_1$ and {\tt class\_name}$_2$ & {\tt C+R} \\
        & & add relevant attributes to  {\tt rel\_name} 
        & {\tt S} \\
        & & annotate  {\tt rel\_name} with relevant [ontologies$|$samples$|$description] & {\tt C+R} \\
        & & introduce attribute description for {\tt rel\_name} 
        & {\tt S} \\  \cline{2-4}
       
        &  \multirow{5}{1.3cm}{{\bf fix}} & rename {\tt rel\_name} & {\tt C+R} \\
        & & enhance cardinality of {\tt rel\_name} & {\tt S} \\
        & & enhance attributes [name$|$type] of {\tt rel\_name} 
        & {\tt S} \\
        & & update the attribute description for {\tt rel\_name} & {\tt R} \\
        & & enhance [ontologies$|$samples] for {\tt rel\_name} & {\tt S} \\ 
        \cline{2-4}
        
& \multirow{1}{1.3cm}{{\bf explain}} &  explain in human-friendly terms the role of {\tt rel\_name} in the schema & {\tt -} \\ \hline  
    \end{tabular}
    \end{footnotesize}
    \caption{Description of main schema editing operations.}
    \label{tab:schemaquery}
\end{table*}

\par{\bf Intelligent \schemalink Workflow.}
A textual description of the target domain $T$ is provided by the curator that needs to develop a new schema from scratch.
The collection $(S)$ of the vector database is queried using the embedding of $T$ to identify previously curated schemas 
in semantically related domains.
Although schemas represent structured knowledge, LinkML schemas are stored as textual artifacts in the repository.
As a result,  a LinkML schema embedding captures thematic and structural features comparable to those present in the textual domain description $T$.
The Top-$k$ most similar schemas are therefore used as contextual examples to guide the LLM in generating a schema.
If the LLM hallucinates and the generated output is not a valid instance of the LinkML language, the system applies syntactic correction routines and, if necessary, re-prompts the LLM until a valid schema is obtained.

\begin{figure*}
    \centering
    \includegraphics[width=.85\linewidth]{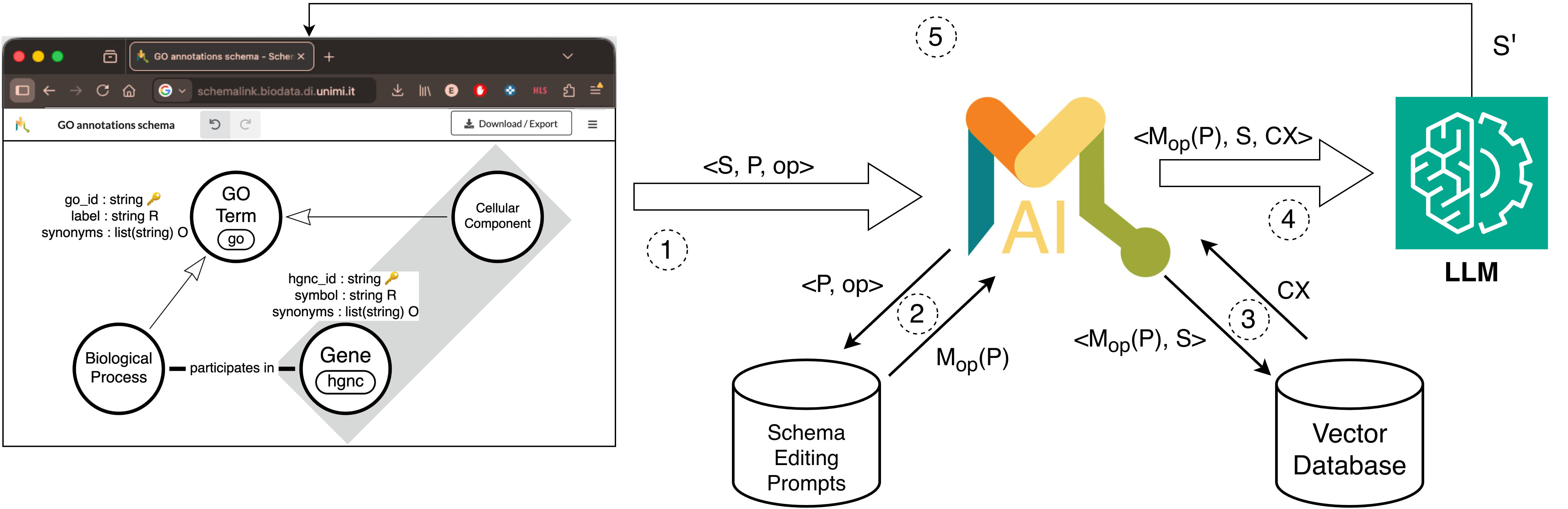}
    \caption{Intelligent SchemaLink workflow.
    }
\label{fig:workflow}
\end{figure*}

The workflow is more articulated when modifying an existing schema. In this case, the canvas already contains a schema $S$, and the curator selects the portion $P \subseteq S$ to be enhanced (e.g. a class, a relationship, or a subgraph).
For example, consider the schema on the left-hand side of Fig.~\ref{fig:workflow}, where the curator selects the classes \texttt{Gene} and \texttt{CellularComponent}. By means of a drop-down menu, the curator identifies the operation $op$ to apply (in 
our example, $op=$\texttt{Add association}). At this point, the intelligent component receives the triplet $\langle S, P, op \rangle$ as input (step 1).

Using $\langle P, op \rangle$, \schemalink retrieves the corresponding prompt template $M_{op}$ from the repository and instantiates it according to the selected portion $P$, yielding $M_{op}(P)$ (step~2). The pair $\langle M_{op}(P), S \rangle$ is embedded to retrieve contextually similar examples $CX$ from the vector database (step~3).
This retrieval strategy differs from a na\"ive RAG setup in which only the 
textual description of $S$ is embedded. 
By jointly embedding $M_{op}(P)$ together with the current schema $S$, retrieval becomes schema- and operation-aware. As a result, the retrieved context $CX$ also depends on the intended operation kind required by the curator. 

\begin{figure}[t]
    \centering
    \includegraphics[width=0.9\linewidth]{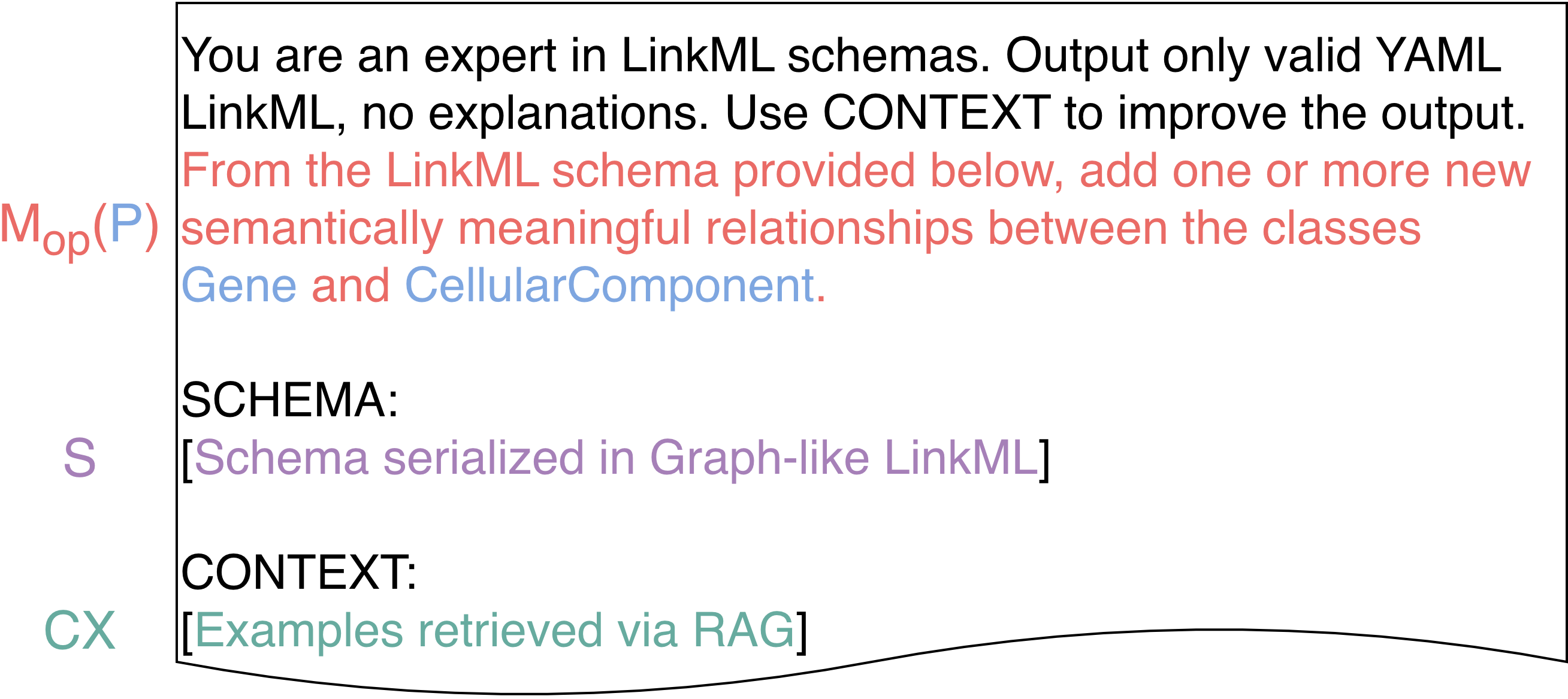}
    \caption{Example of LLM prompt for generating a new association.}
    \label{fig:addassociationprompt}
\end{figure}

Finally, \schemalink combines $\langle M_{op}(P), S, C_X \rangle$ to construct a LLM prompt for the generation of an updated schema $S'$ (step~4). Fig.~\ref{fig:addassociationprompt} illustrates the structure of the prompt in the running example. The generated schema $S'$ undergoes automatic syntactic validation to address potential hallucinations as discussed above. After validation, $S'$ is rendered on the canvas for user inspection (step~5). The curator may accept the proposed schema or further refine it either manually or by applying additional schema editing operations. In our example, the system proposes a new association named {\tt located in} between {\tt Gene} and {\tt CellularComponent}. 
A complete example illustrating how the \schemalink intelligent component can be used to enhance our running example in a transcriptomics context is provided in the Supplementary Materials.

\subsection{Representation of LinkML Structural Forms}\label{sec:importexport}

The graphical artifacts introduced in the previous sections support the generation and serialization of different LinkML {\em structural forms}~\citep{structuralforms} that allow schema designers to choose the most appropriate representation for their data and technical constraints.

Selecting a structural form depends on the complexity of the domain, the nature of the relationships, interoperability requirements, query performance, and the intended data infrastructure (e.g. document-oriented, property/RDF graphs).

In the current implementation, \schemalink supports the two most commonly adopted structural forms: tree-like and graph-like representations. Schemas expressed in either form can be imported into \schemalink to generate the corresponding graphical representation, and vice-versa, schemas created within \schemalink can be serialized into these forms. During import, schemas undergo syntactic validation using standard LinkML community tools to ensure correctness.

The tree-like form organizes data in a nested manner, where associations are typically embedded within classes as attributes. This style is well-suited for hierarchical or document-oriented data models where nested groupings play a crucial role.
Classes may {\em reference} classes or can be nested using the {\em inlined} clause. Class inheritance is supported
via the \texttt{is\_a} clause, enabling reuse of attributes and annotations.
This form is thus ideal to model complex, interrelated data elements that are not easily captured in a flat structure (e.g. organizational structures or product catalogs). Data conforming to this form are usually serialized in JSON-like or document-based databases (e.g. MongoDB~\citep{mongo}).

\begin{figure*}
    \centering
    \includegraphics[width=0.95\linewidth]{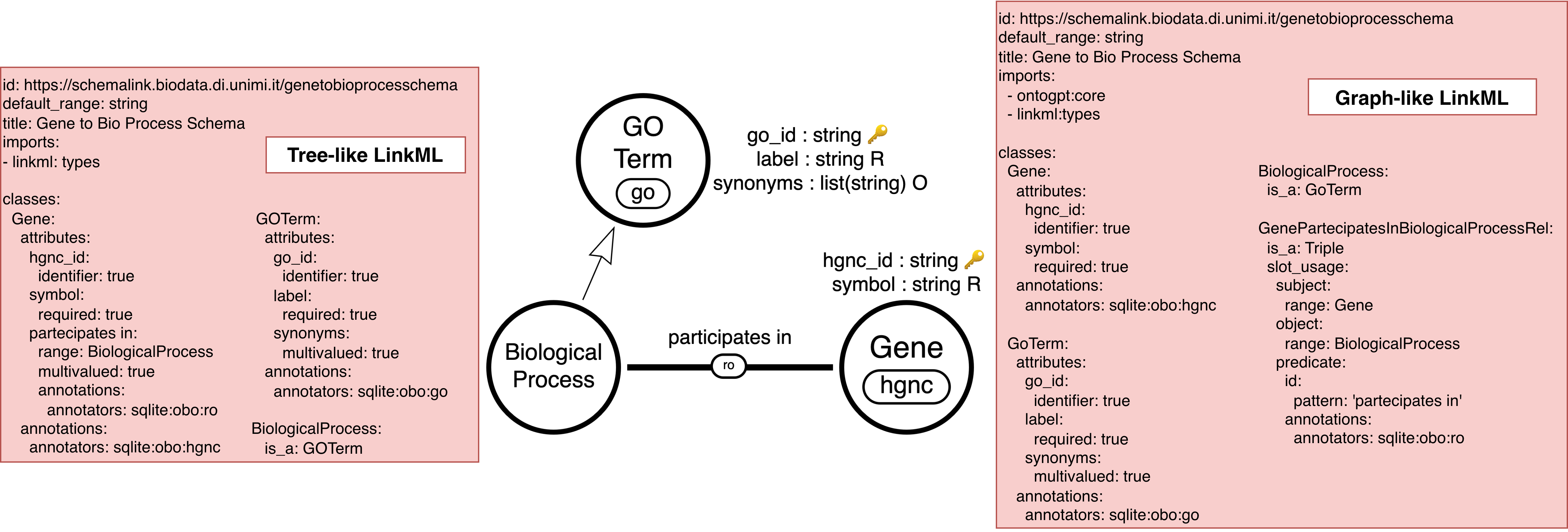}
    \caption{Tree- and graph-like LinkML structural forms.}
    \label{fig:mapping}
\end{figure*}

\begin{example}
Consider the schema excerpt 
in Fig.~\ref{fig:mapping}. 
In the tree-like form (left panel), the association {\tt Gene-participates in-BiologicalProcess} is defined as an attribute 
of the class \texttt{Gene}. 
Its {\em many-to-many} cardinality is represented using the \texttt{multivalued} clause. 
The class \texttt{BiologicalProcess} is declared as a subclass of \texttt{GOTerm} via the \texttt{is\_a} clause. 
\end{example}

\vspace*{-9pt}

The graph-like form represents entities as nodes and associations as edges, 
with properties attached directly to nodes and edges. Inheritance is represented as in the tree-like form.
The graph-like form is well-suited for detailed, attribute-rich models. Data conforming with this form are usually represented as property graphs, e.g. via Neo4j~\citep{neo4j}) or the RDF-star~\citep{rdfstar} data model. 
For this form, \schemalink uses  the OntoGPT~\citep{spires} core classes (mainly {\tt NamedEntity} and {\tt Triple}) and the property graph schema paradigm~\citep{pgform}.

\begin{example}
The schema in Fig.~\ref{fig:mapping} can also be represented in the graph-like form (right panel). 
Here, the association between \texttt{Gene} and \texttt{BiologicalProcess} is explicitly modeled as a class 
with attributes defining the {\tt subject}, {\tt predicate} (i.e. the relationship type), and {\tt object} of the relationship.
\end{example}

This dual representation of schemas has the positive side effect that schemas represented through the tree-structured form can be translated into the graph-structured form and vice-versa through a heuristic algorithm that we have developed~\citep{paoloalgo}.

\section{Implementation}

The ChromaDB~\citep{chromadb} vector database
was used for indexing schemas via the OpenAI {\em text-embedding-3-large} model.
For feeding the vector database we used
a repository of 60 expert‑curated LinkML schemas from the OntoGPT project (available at
\url{https://github.com/monarch-initiative/ontogpt/tree/main/src/ontogpt/templates}).
To reduce variability arising from heterogeneous schema formats and LinkML dialects, we generated a graph-like schema form through the SchemaLink import/export routines, followed by manual quality control. Starting from the collection $(S)$, we generated the 
downstream collections. Specifically, collection $(C)$ contains 157 classes,  collection $(R)$ contains 127 relationships, and 
collection $(C\!\!+\!\!R)$ contains the same number of relationships, but enriched with the class specifications.
The LLM used for the intelligent component is {\em gpt‑4o‑mini} because it provides a good balance between responsiveness and output quality. We set {\em temperature} to $0$ and the number of retrieved examples via RAG to $10$ items for each intelligent operation.

Users can register with SchemaLink to access its intelligent functionalities. Registration is required because the RAG-based component relies on GPT models, whose usage incurs operational costs. Registered curators can introduce their schemas to enrich the vector database. This community-driven enrichment process improves retrieval quality and broadens the semantic coverage of the RAG component.
The application and the schema collection are released as open source. This allows users to inspect indexed schemas, run SchemaLink locally, and configure the system with their own API keys (details at: \url{https://anacletolab.github.io/schemalink-docs}).

To evaluate the \schemalink intelligent component (both the generation of schemas from scratch and refined schemas obtained by the application of the schema editing operations), we considered four state-of-the-art conversational LLM systems -- ChatGPT (OpenAI), DeepSeek, Claude (Anthropic), and Gemini (Google) -- alongside ten
expert human curators. 
Although human curators serve as the primary reference standard, LLMs provide more scalable estimates. 

The LLM evaluation followed the LLM-as-a-judge paradigm~\citep{llmjudge}.
In the LLM‑as‑a‑judge setting, more capable models (in terms of parameter count and reasoning ability) are employed to evaluate the outputs of the lighter‑weight model used as \schemalink engine (i.e. {\em gpt‑4o‑mini}). 
To better approximate realistic usage scenarios in which schema proposals are iteratively refined, we decided to exploit the official web-based conversational interfaces of these systems because they preserve interaction history (the use of the stateless API calls would have required reconstruction of dialog context at each step).
Moreover, using independent LLM judges improves fairness and reduces vendor‑specific bias (e.g. avoiding the situation in which only ChatGPT is used to evaluate a schema produced by a GPT‑based engine). Prompts for interacting with LLM-judges are provided in the Supplementary Materials.

\section{Discussion}

Several experiments have been conducted to evaluate, on the one hand, the quality of the schema generated by the intelligent component and, on the other hand, its efficiency and applicability in a web environment.  Specifically, we describe the experiments for evaluating the quality of the schemas generated from scratch and the quality of the enhanced schemas obtained through the intelligent operations. Moreover, we have evaluated the impact of different vector database collections on the RAG-based retrieval strategy.
Finally, we discuss the execution times of the intelligent operations.

\begin{table}[b]
\centering
\scriptsize
\begin{tabular}{|l|*{5}{c|}}
\hline
& \textbf{Disease} & \textbf{Drug} & \textbf{Protein} & \textbf{Pathway} & \textbf{RNA} \\
\hline
\textbf{ChatGPT} & 4 & 4 & 3 & 4 & 3 \\
\hline
\textbf{DeepSeek} & 3 & 4 & 3 & 3 & 3 \\
\hline
\textbf{Claude} & 3 & 4 & 3 & 4 & 3 \\
\hline
\textbf{Gemini} & 4 & 4 & 3 & 4 & 3 \\
\hline
\textbf{Curators} & 
2.4$\pm$.5 & 
3.7$\pm$.5 & 
3.0$\pm$.8 & 
3.7$\pm$.5 & 
4.5$\pm$.5
\\
\hline
\end{tabular}
\caption{Evaluation of schemas generated from scratch.}
\label{tab:generateeval}
\end{table}

\begin{figure}[b]
    \centering
    \includegraphics[width=\linewidth]{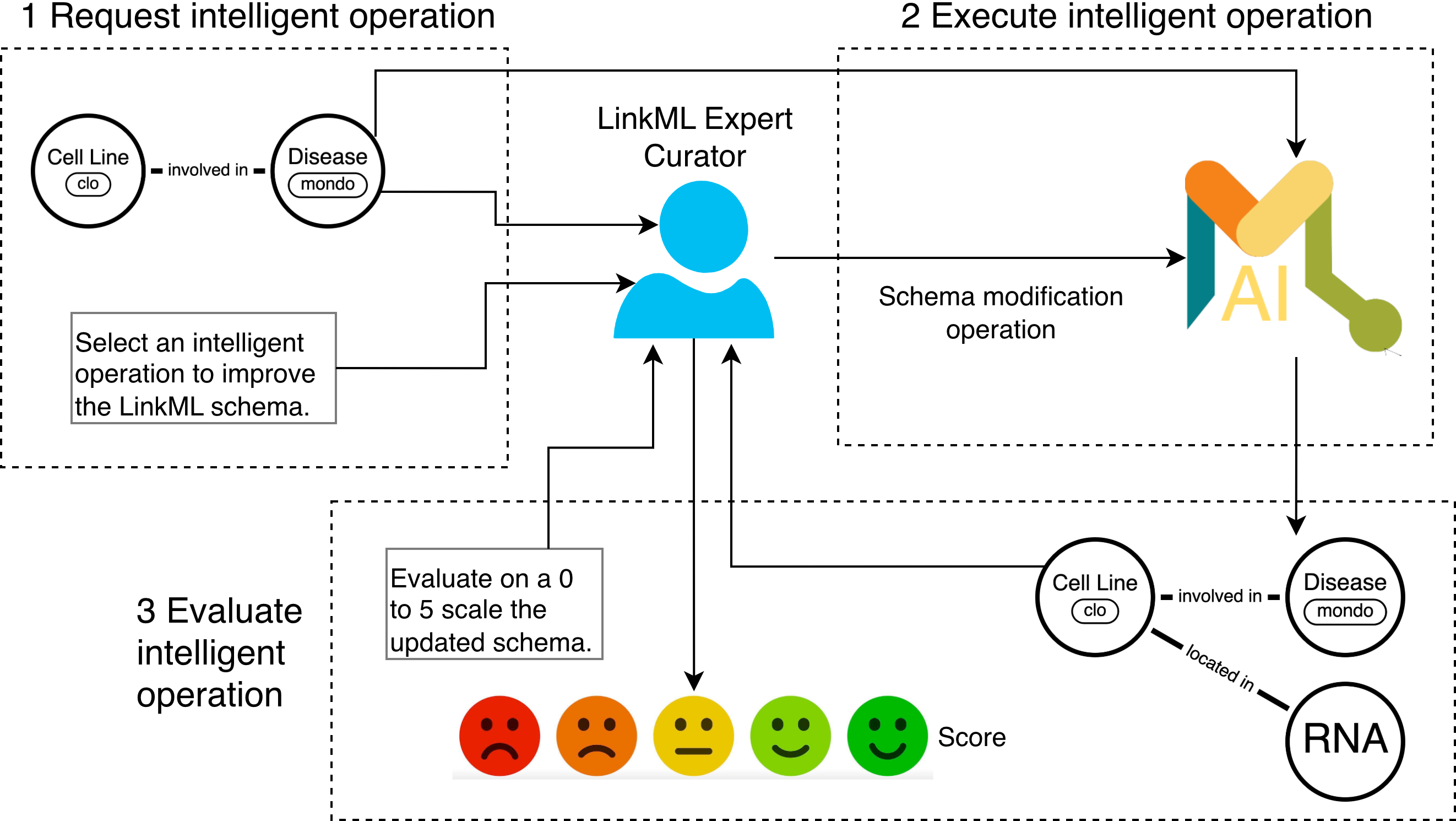}
    \caption{Evaluation pipeline for \schemalink intelligent operations.}
    \label{fig:evalpipeline}
\end{figure}

\subsection{Quality Evaluation of Schemas generated from Scratch}

To evaluate the quality of the schema developed from scratch, we defined five case studies in collaboration with biomedical LinkML curators. Each case corresponds to a textual description of a domain of interest, and \schemalink is used to generate an initial schema.
Then, the expert is asked to assign a discrete score from 1 to 5 to the generated schema according to the following guidelines: 1 = schema unintelligible or not applicable to the domain; 2 = largely incorrect with major inconsistencies; 3 = broadly usable but requiring several revisions; 4 = correct and coherent with only minor refinements needed; 5 = fully correct and well‑suited to the domain. Adopting a five-value scale for schema quality aligns with research practices for subjective evaluation~\citep{likertscale}.

Use cases correspond to schemas that model: $(i)$ entities involved in diseases (\texttt{Disease}); $(ii)$ drugs and their mechanisms of action (\texttt{Drug}); $(iii)$ proteins, including functional domains and post‑translational modifications (\texttt{Protein}); $(iv)$ biological pathways and their participating entities (genes, RNAs, proteins, metabolites; \texttt{Pathway}); and $(v)$ relations among RNA molecules, including subtypes such as ncRNAs and mRNAs, and their further specializations such as miRNAs and siRNAs (\texttt{RNA}).
Prompts issued to the intelligent component of \schemalink for generating schemas from scratch in these domains are provided in the Supplementary Materials.

Table~\ref{tab:generateeval} reports the scores assigned to the {\em Generate} operation across the five use cases. For domain experts, we also include the standard deviation across the ten evaluations.
Scores provided by LLM-judges are consistent across models, and $9/20$ ratings (45\%) are greater than or equal to 4, which (according to our protocol) indicates that the generated schema is coherent and needs minor refinements. {\tt Protein} and {\tt RNA}, which require more specialized subclassing, are typically judged semantically correct but incomplete. We also observe that ChatGPT and Gemini assign slightly higher average scores (both $3.6$) than Claude ($3.4$) and DeepSeek ($3.2$); this may reflect family alignment with the GPT‑based engine used in \schemalink ({\em gpt‑4o‑mini}), which was likely trained on the same corpus of data as the ChatGPT model.
Domain expert evaluations are overall aligned with those of LLM-judges: on average, $4/5$ use cases receive scores above 3, with {\tt RNA} being considered the most complete ($4.5$). The reported standard deviations are generally low since expert ratings typically differ by at most one point.
Qualitative feedback from domain experts consistently highlights that classes and hierarchies are well modeled, while relationship and attribute types are often the least semantically refined components.

\subsection{Quality Evaluation of Schema Editing Operations} 

To evaluate the quality of the schema editing operations, we have adopted the pipeline described in Fig.~\ref{fig:evalpipeline}. 
Starting from the schema generated in the previous step, the curator is asked to choose an intelligent operation (step~1 in the figure).
Then, the SchemaLink intelligent component proposes an updated schema to the user (step~2). 
Finally, the curator is asked to score on a 0–5 scale the relevance of the obtained schema (step~3). 
The guideline for assigning the score is the following: 
0 = failure to display any changes; 1 = failure to accomplish the requested change; 2 = substantial manual repair needed; 3 = the change is achieved with minor issues; 4 = correct change with negligible adjustments; and, 5 = fully correct change with no adjustments.
This procedure is repeated six times per schema (i.e. six intelligent operations are evaluated) to capture performance across a representative editing sequence.

Fig.~\ref{fig:piechart} summarizes the average scores obtained for each case and judge. Out of 420 evaluations
, 363 (86.4\%) 
achieve a score greater than or equal to 3, and 279 (66.4\%)
achieve a score greater than or equal to 4. This means the vast majority of the editing operations are semantically sound.
It is worth noting that 9 evaluations 
referring to operations that modify attribute types were assigned a score of 0 by LLM-judges. We inspected such cases and noticed that the target already had an appropriate type, meaning no change was required (i.e. the LLM hallucinated). We re-prompted the LLMs, and they confirmed that these low scores reflected an inappropriate choice of operation rather than a failure of the intelligent component.

\begin{figure}[t]
  \centering
  \begin{tikzpicture}

\node at (0,3.4) {\bfseries Editing Operation Evaluation};
\pie[
    radius=3,
    color={red!40, orange!40, yellow!50, teal!40, blue!20, violet!40},
    hide number
]{
    35.0/{},
    31.4/{},
    20.0/{},
    10.2/{},
    1.2/{},
    2.1/{}
}

\node[align=center,font=\bfseries] at (0.85,1.3) {5\\[2pt]\normalfont 35.0\%\\\normalfont (147)};
\node[align=center,font=\bfseries] at (-1.7,0) {4\\[2pt]\normalfont 31.4\%\\\normalfont (132)};
\node[align=center,font=\bfseries] at (0.15,-1.7) {3\\[2pt]\normalfont 20.0\%\\\normalfont (84)};
\node[align=center,font=\bfseries] at (3.45,-0.20) {0 \normalfont (9)};
\node[align=center,font=\bfseries] at (2.625,-0.1735) {\normalfont 2.1\%};
\node[align=center,font=\bfseries] at (2.0,-1.15) {2\\[2pt]\normalfont 10.2\%\\\normalfont (43)};
\node[anchor=west] at (2.9,-0.6) {{\bf 1} 1.2\% (5)};

\end{tikzpicture}

  \vspace{0.125cm}
  \scriptsize
  \begin{tabular}{|l|*{5}{c|}}
  \hline
  & \textbf{Disease} & \textbf{Drug} & \textbf{Protein} & \textbf{Pathway} & \textbf{RNA} \\
  \hline
  \textbf{ChatGPT} & 4.3 & 4.3 & 4.5 & 3.5 & 4.0 \\
  \hline
  \textbf{DeepSeek} & 3.7 & 4.5 & 3.5 & 4.0 & 3.3 \\
  \hline
  \textbf{Claude} & 3.8 & 3.3 & 3.5 & 2.7 & 2.7 \\
  \hline
  \textbf{Gemini} & 3.3 & 3.7 & 3.2 & 2.3 & 3.5 \\
  \hline
  \textbf{Curators} & 4.0
  & 3.7
  & 4.2
  & 4.0
  & 3.7
  \\
  \hline
  \end{tabular}

  \caption{Average scores for intelligent editing operations.
  }
  \label{fig:piechart}
\end{figure}

Overall, \emph{Add} operations achieve the highest performance among all categories, with an average score of around 4, and they were  judged more than 200 times by human curator and more than 20 times by each LLM-judge. A plausible explanation is that these operations benefit from a larger set of relevant examples stored in the RAG collections, which enhances contextual grounding during generation. Feedback from expert curators highlights that the generated classes and relationships are generally appropriate and well-formed, requiring only minor adjustments (e.g. refinement of attribute types).
\emph{Fix} operations--which require contextual reasoning over schema structure and constraints--achieve an overall average score of 3.3. They were invoked 70 times by human curators (mean score 3.7), 8 times by ChatGPT (mean score 3.0), Gemini (2.2), and Claude (1.7), and 10 times by DeepSeek (3.0).
Feedback from curators suggests that fixes applied to subgraphs are particularly effective and appreciated, as they enable one-shot refinements of multiple elements. However, more fine-grained constraints on attributes (e.g. enforcing that a \emph{p-value} must be lower than 0.05) are sometimes under-specified and could be improved.
The \emph{Reify} category was invoked ten times by expert curators (average score 4.5) and only once by LLMs (Gemini), which assigned a score of 3 to the resulting schema. Curators noted that reifications capture the intended structural transformation; however, they also expected the inclusion of specialized attributes or ontology for the resulting subclasses.

\begin{figure}[t]
    \centering
    \includegraphics[width=\linewidth]{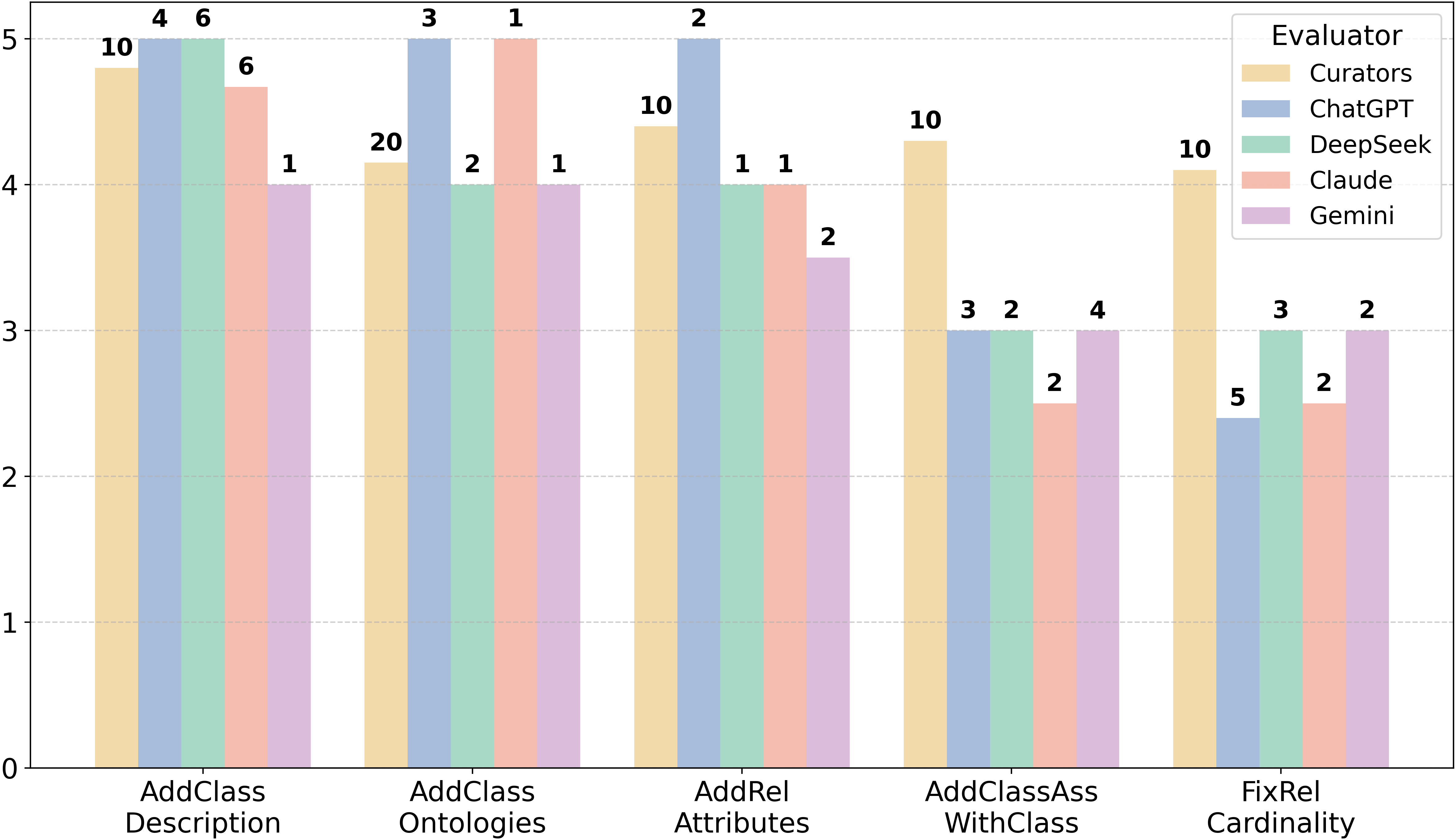}    
    \caption{Top‑5 operations by average score.}
    \label{fig:AvgEvalPerLLM}
\end{figure}

Fig.~\ref{fig:AvgEvalPerLLM} details the five highest-performing intelligent operations among those invoked at least once by each evaluator. Bar height represents the average score assigned by each evaluator, while the number displayed above each bar indicates how many times the corresponding operation was selected and evaluated.
\emph{Add} operations are both frequently invoked and highly rated. In particular, \texttt{AddClassDescription} was invoked 27 times and achieved an average score of 4.7. Similarly, \texttt{AddClassOntologies} and \texttt{AddRelAttributes} received consistently strong evaluations across judges.
The assessments provided by expert curators are closely aligned with those of LLM-judges for these operations.
Among {\em Fix} operations, \texttt{FixRelCardinality} achieves a mean score of 3 across 22 invocations, suggesting that the intelligent component can support schema refinements. Notably, \texttt{FixRelCardinality} and \texttt{AddClassAssociatedWithClass} receive higher evaluations from curators (exceeding LLM scores by more than one point on average). This suggests human evaluators better appreciate structurally sound and context-dependent refinements, particularly when modeling complex relationships.

Finally, we asked expert curators to answer three additional questions assessing $(i)$ the overall quality of the final schema, $(ii)$ the effectiveness of the enhancement process from the initial to the refined schema, and $(iii)$ the level of effort that would have been required to perform the same refinement manually, without intelligent assistance. 
On a 1--5 scale, the average score is around 4 across all questions. Qualitative feedback from curators highlights that relationships are often less accurate than classes; however, generated schemas are semantically acceptable and solid starting points for further refinement.

We also asked curators to evaluate whether the final schema was structurally coherent and meaningful (\emph{Yes}/\emph{Partially}/\emph{No}). In 28 cases (56\%), the answer was \emph{Yes}, further supporting the overall quality of the system.

\subsection{Impact of ``Custom'' RAG Collections}

We tested whether the organization in four collections of the vector database improves the quality of intelligent schema editing compared with a single collection containing only entire LinkML schemas. We evaluate each use case four times by considering the four collections ($C$, $R$, $C\!\!+\!\!R$, $S$) as example sources for the RAG component. We relied only on the LLM-as-a-Judge technique to scale this analysis because a complete human evaluation of all configurations would be prohibitively expensive (we consider the 4 LLM-judges, 5 running use cases, 4 retrieval collections, and 6 operations resulting in $480$ interactions with the system).

\begin{figure}
    \centering
    \includegraphics[width=\linewidth]{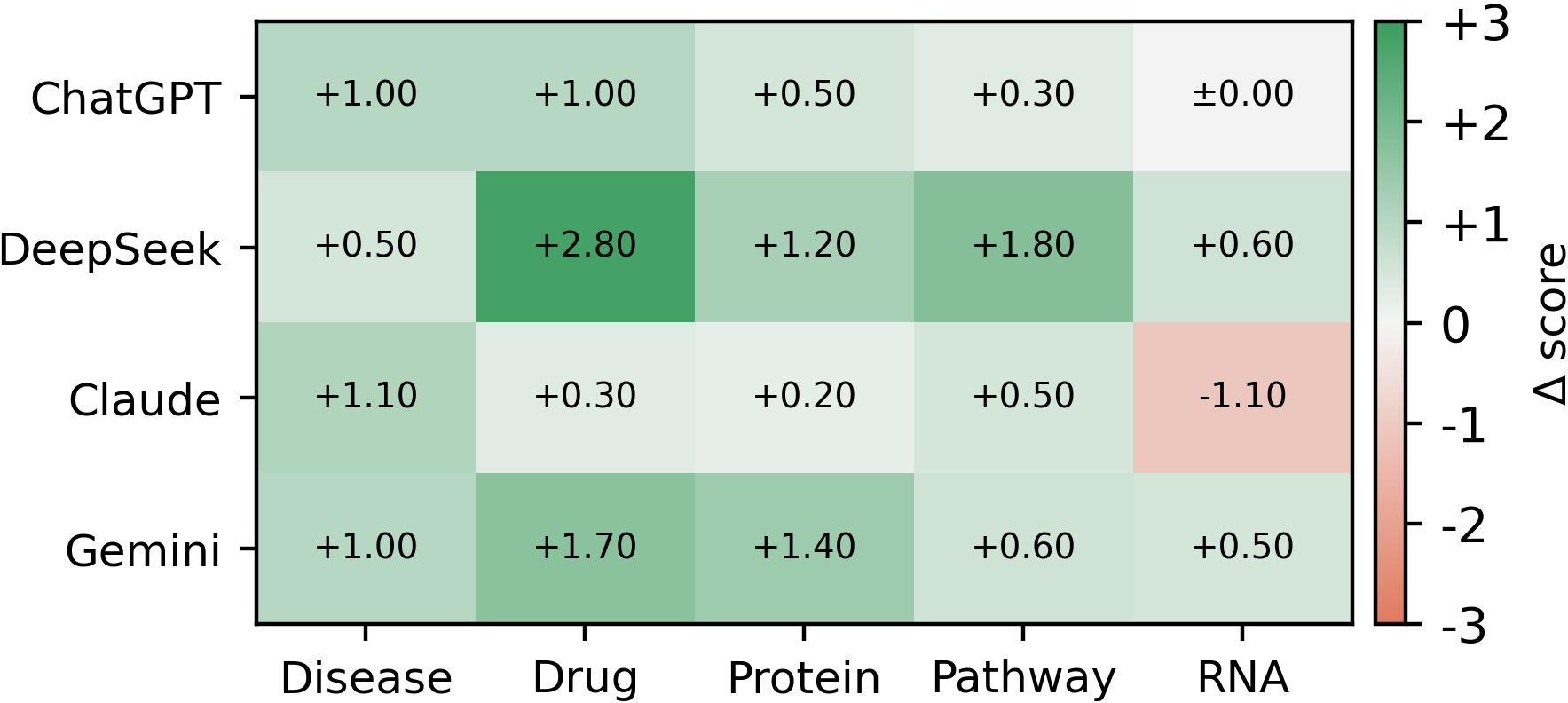}
    \caption{Impact of our custom RAG collections wrt.\ $(S)$.}
    \label{fig:delta}
\end{figure}

Fig.~\ref{fig:delta} reports the average score differences obtained by using our custom collections depending on the editing operation with respect to the baseline $(S)$ strategy.
Similar trends are observed across the other collections; the corresponding heatmaps and detailed statistics are provided in Supplementary Materials.
Overall, our strategy improves generation quality. Gains reach up to +2.8 points (+56\%) in the \emph{Drug} use case under DeepSeek, with an average improvement of +0.8 points (+16\%) across all models and domains. Average improvements per LLM-judge are +0.56 (ChatGPT), +1.38 (DeepSeek), +0.20 (Claude), and +1.04 (Gemini), indicating that the benefit of targeted retrieval is robust across judges. Only one decrease is observed, for the \emph{RNA} case under Claude (-1.10). 

\subsection{Time Performance}\label{sec:runningtimes}
To evaluate time performance, we executed the full set of 43 intelligent operations three times each on three different schemas using a commodity laptop connected to the SchemaLink web application via the {\it Google Chrome} browser (release {\it 138}, its developer tool was used to assess time latencies). Table~\ref{tab:deviazione_media_operazioni_per_tipo} reports the average end-to-end latency and standard deviations, grouped by both operation category and target type (class, relationship, or subgraph). Overall, the measured latencies are often below 15~seconds. Moreover, the results indicate that {\em Generate}, {\em Reify}, and operations acting on a subgraph require noticeably more time. This is expected, as these operations either create new schema components from scratch or manipulate sub‑schemas that typically involve multiple classes and relationships.

Latency also benefited from custom RAG collections. 
When relying only on the $(S)$ collection, complex operations could take several minutes due to the large context window. 
Retrieving targeted schema portions reduces the context size and leads to faster inference. 
This is consistent with prior findings showing that smaller context windows improve LLM inference efficiency and latency~\citep{contextwindowsize1,contextwindowsize2,contextsize3}.

\begin{table}[ht]
\begin{minipage}[t]{0.4\columnwidth}
\footnotesize
\begin{tabular}{l|c}
\hline
\textbf{Operation} & \textbf{Mean$\pm$Std.(s)} \\\hline\hline
Generate & $9.68\pm1.99$ \\
Add      & $5.05\pm1.84$ \\
Fix      & $6.47\pm4.05$ \\
Reify    & $14.61\pm4.87$ \\
\hline
\end{tabular}
\end{minipage}
\ \hspace*{24pt}\
\begin{minipage}[t]{0.4\columnwidth}
\footnotesize
\begin{tabular}{l|c}
\hline
\textbf{Target} & \textbf{Mean$\pm$Std.(s)} \\\hline\hline
Class     & $5.71\pm3.81$ \\
Relation  & $5.48\pm3.52$ \\
Subgraph  & $7.71\pm3.05$ \\
\hline
\end{tabular}
\end{minipage}

\caption{Average execution times by operation category and target type.}
\label{tab:deviazione_media_operazioni_per_tipo}
\end{table}

\subsection{Concluding Remarks}

The experimental campaign demonstrates the effectiveness of our approach.
Experts' evaluations yield high scores and indicate that the system is effective in supporting schema creation and enrichment tasks. At the same time, they highlight important directions for future improvements, as operations involving complex relationships (i.e. those requiring structural reasoning and constraint handling) remain more challenging.
End-to-end latencies are typically below 15 seconds, even for complex operations such as schema generation or reification.

The graphical representation of schemas adopted in this paper can be seen as a ``bridge'' across the several LinkML structural forms. Translation algorithms have been defined to move from one representation to the others. Thus, users can adopt the most convenient form for their domains. At the current stage, the platform supports tree-like and graph-like forms, but further forms (e.g. tabular and relational) can be included.

Future developments will focus on extending supported LinkML structural forms and on broadening import/export capabilities to other schema modeling paradigms such as RDF Schema~\citep{rdfschema}, OWL~\citep{owl}, SHACL~\citep{shacl}, and PG-schema~\citep{pgschema}. Furthermore, we plan to enhance schema modeling by allowing users to specify additional structural and semantic constraints that go beyond syntactic validation. For instance, users could define temporal consistency constraints (e.g. an attribute {\em year} associated with a person cannot refer to a date later than the person's death) and ontology-based constraints (e.g. instances of a class {\tt Gene} associated with a {\tt Species} must comply with the identifiers for that species). These constraints can be integrated into the intelligent component so that schema semantic inconsistencies are reduced during generation and editing.


\bibliographystyle{plainnat}
\balance
\bibliography{reference}

\end{document}


\maketitle

\setcounter{table}{0}
\renewcommand{\thetable}{S\arabic{table}}
\renewcommand{\tablename}{Supplementary Table}

\setcounter{figure}{0}
\renewcommand{\thefigure}{S\arabic{figure}}
\renewcommand{\figurename}{Supplementary Fig.}

\renewcommand{\thelstlisting}{S\arabic{lstlisting}}
\renewcommand{\lstlistingname}{Supplementary Listing}

\section{Vector Database Indexing Filter}\label{app:vectordb}

Schemas are indexed in the vector database as textual representations, meaning that classes and relationships are embedded as strings. 
Since different schemas may contain syntactically different but semantically equivalent entities (e.g., classes named {\tt Gene} and {\tt Genes}), we apply a near-duplicate filtering procedure before indexing items into each collection.

For any pair of classes or relationships $c_i$ and $c_j$, we compute a string-based distance over their names:
$d_{name}(c_i,c_j) = \mathrm{Lev}\big(\mathrm{name}(c_i), \mathrm{name}(c_j)\big),$
where $\mathrm{Lev}$ denotes the Levenshtein distance.
If $d_{name}(c_i,c_j) < 0.2$, the two elements are considered potential duplicates. 
In this case, additional comparisons are performed on their descriptions, attribute definitions, and ontology annotations.
If
$d_{desc}(c_i,c_j) < 0.3\text{ }\And\text{ }
d_{attr}(c_i,c_j) < 0.3\text{ }\And\text{ }
d_{ann}(c_i,c_j) < 0.3$,
the two entities are considered duplicates, and only the element with the largest total character length is indexed in the database.

\section{\schemalink at Work}

Suppose a biomedical curator is interested in organizing data about functional genomics experiments that were performed across research groups. 
A common scenario involves transcriptomic studies (e.g. RNA-seq experiments) in which gene expression is measured under specific experimental conditions, and genes are subsequently annotated with functional information. 
In such contexts, it is essential to:~$(i)$~uniquely identify genes to avoid duplicate or ambiguous entries;~$(ii)$~characterize the experimental setting (treatment, biological sample, perturbation, etc.);
and~$(iii)$~annotate genes with GO terms.

Although LinkML can be profitably leveraged to design the schema for this domain and include usual entities, relationships, and properties that applications need, its manual writing is time-consuming, and domain curators should inspect previously developed schemas to generate an accurate new one. 

By means of \schemalink, curators can start from a simple textual description of the domain and obtain an initial schema that takes advantage of the schema already loaded in the vector database. 
For example, with the following prompt
\emph{``Generate a LinkML schema involving experiments that measure the expression of genes, and include the cellular components in which genes are located. Cellular components can be modeled as specializations of Gene Ontology terms.''}, the initial schema in Supplementary Fig.~\ref{fig:schemaintermediate} is generated.
Its structure captures key requirements: genes are uniquely identified via HGNC identifiers, experiments are explicitly represented as first-class entities, and annotations are structured according to the desired GO categories.

Although genes are already linked to \texttt{CellularComponent}, functional genomics studies, like gene enrichment, typically require broader functional annotations, since genes are also interpreted in terms of the processes, functions, and pathways in which they participate.
To support this extended functional characterization, through the drop-down menu for the class \texttt{Gene}, the curator invokes the operation {\em Add class associated with}. 
\schemalink introduces the new class named \texttt{BiologicalProcess}, modeled as a specialization of \texttt{GOTerm}, together with the corresponding association {\tt Gene-participates in-BiologicalProcess}.
After adding a few attributes to the newly introduced elements, the schema evolves into the one shown in Fig.~1.

Furthermore, gene expression experiments are frequently performed in different organisms (e.g. human, mouse, Drosophila), and results depend on the species under investigation. For example, differential expression patterns observed in a mouse model may need to be compared with their human orthologs in comparative and translational studies.
For this reason, to explicitly represent the organisms involved in genomic experiments, the curator manually introduces the class  \texttt{Species}.
Then, through {\tt Add} operations, the intelligent component enriches this class by proposing the ontology NCBITaxon, a description (\emph{``A taxonomic species, representing a group of organisms capable of interbreeding.''}), and a set of candidate attributes.
The set of attributes includes:
$(i)$ \texttt{name} (required), described as ``the common name of the species'';
$(ii)$ \texttt{scientific\_name} (required), representing ``the scientific (Latin) name of the species'';
$(iii)$~\texttt{habitat} (optional), ``the natural environment in which the species lives'';
$(iv)$ \texttt{conservation\_status} (optional), ``the extinction risk category''; and
$(v)$ \texttt{synonyms} (optional), ``a list of alternative names for the species''.

Further enhancements can be applied to the schema either manually or through the intelligent component. For example, basic {\tt string} types of attributes can be 
enhanced to more specific types (such as a controlled vocabulary for the attribute \texttt{conservation\_status}) or instance-level examples that can annotate classes or associations.
Further associations can also be suggested by the \schemalink engine (like {\tt Experiment-conducted on-Species} and {\tt Experiment-has-Species}). The curator can inspect alternatives and retain only the semantically appropriate ones (e.g. discarding \emph{has} because it is too generic).

More advanced modeling patterns are also supported. For example, a recursive relationship on 
\texttt{Gene} can be introduced to represent homologous genes. After its creation, the {\tt Fix} operation can be applied to improve 
its name. The system proposes \emph{1 to 1 homologous to}, which is aligned with RO.

At the end of this iterative refinement process, the schema in Supplementary Fig.~\ref{fig:finalschema} is obtained.
The classes \texttt{Experiment} and \texttt{Species} are introduced, with attributes and ontology annotations.
Moreover, \texttt{Species} is embedded in \texttt{Gene} via a required attribute \texttt{species}, ensuring that each gene instance is associated with an organism.
The relationships \texttt{Experiment-conducted on-Species} and \texttt{Gene-1 to 1 homologous to-Gene} are proposed.
Since it is grounded in established bio-ontologies, it provides a framework for integrating in-house experiments and data retrieved from centralized repositories.

\begin{figure}[ht]
    \centering
    \includegraphics[width=0.7\linewidth]{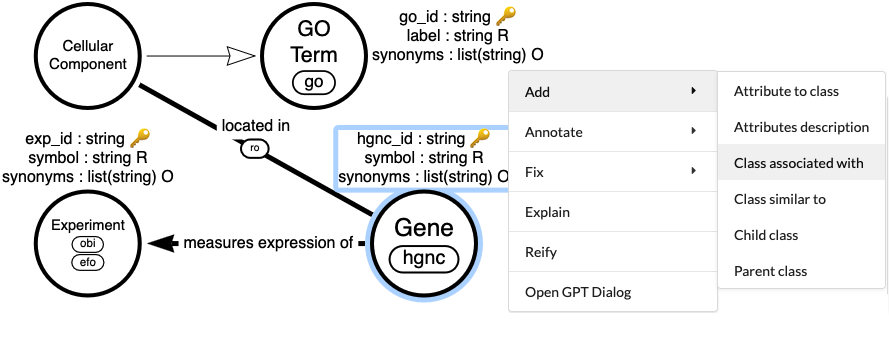}
    \caption{Intermediate schema and menu with intelligent operations.}
    \label{fig:schemaintermediate}
\end{figure}

\begin{figure}[ht]
    \centering
    \includegraphics[width=0.85\linewidth]{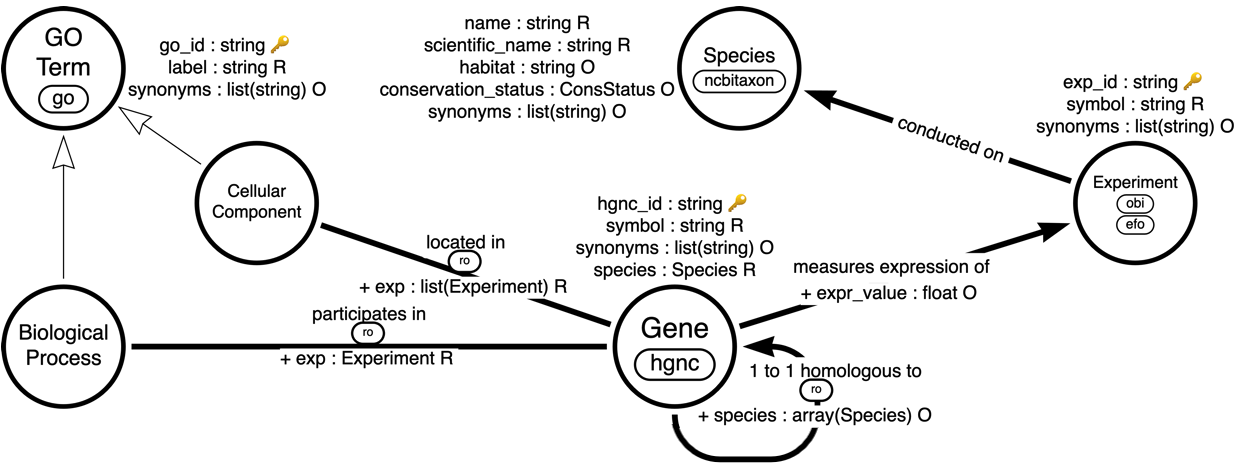}
    \caption{Graphical representation of the schema obtained after editing operations.}
    \label{fig:finalschema}
\end{figure}

\clearpage

\section{LLM-as-a-Judge Prompts}\label{app:llmjudgeprompt}

This supplementary section reports the prompts used within the LLM-as-a-Judge evaluation framework. 
The evaluation protocol consists of three stages, each associated with a specific prompt:
$(i)$ evaluating the initial schema generated via \schemalink (Supplementary Listing~\ref{lst:generateevalprompt}, shown for the {\tt Disease} use case);
$(ii)$ selecting an intelligent editing operation to improve the schema (Supplementary Listing~\ref{lst:requestprompt}); and~$(iii)$~evaluating the updated schema after applying the operation (Supplementary Listing~\ref{lst:evaluateOperation}).

\begin{lstlisting}[language={}, 
                   basicstyle=\scriptsize\ttfamily,
                   escapeinside={(*}{*)},
                   caption={Prompt to evaluate the intial schema.},
                   label={lst:generateevalprompt},
                   xrightmargin=-12pt]
You are an expert LinkML schema designer working in the biological and biomedical fields. Note that in your domain of
expertise, LinkML relations are represented using the Triple core class and their predicates are represented using the
RelationshipType core class.

Rate my LinkML schema that is described as "a LinkML schema involving diseases" from 1 to 5, where:
1: Poorly written and poorly coherent with the description I gave you
2: Contains significant issues in structure, consistency, but a partial attempt at schema design is visible.
3: Adequately written, but lacking clarity, completeness according to your knowledge.
4: Well-written, mostly coherent with the intended schema description, with only minor issues or inconsistencies.
5: Excellent, fully coherent with the schema description, structurally consistent, and clearly written.

Schema:
  {LinkML schema}
\end{lstlisting}

\begin{lstlisting}[language={}, 
                   basicstyle=\scriptsize\ttfamily,
                   escapeinside={(*}{*)},
                   caption={Prompt to request an intelligent operation.},
                   label={lst:requestprompt},
                   xrightmargin=-12pt]
You are an expert LinkML schema designer working in the biological and biomedical fields. Note that in your domain of
expertise, LinkML relations are represented using the Triple core class and their predicates are represented using the
RelationshipType core class. I am asking you as an expert to choose an operation to improve the LinkML schema proposed
below. Choose an operation from the following list and specify the class, relationship, or schema portion on which you
would like to
execute the operation.

Schema:
  {LinkML schema} 
  
List:
  {List of intelligent operations with their descriptions}
\end{lstlisting}

\begin{lstlisting}[language={}, 
                   basicstyle=\scriptsize\ttfamily,
                   escapeinside={(*}{*)},
                   caption={Prompt to evaluate the intelligent operation.},
                   label={lst:evaluateOperation},
                   xrightmargin=-12pt]
I have updated the LinkML schema according to the operation you chose. Rate my LinkML schema from 0 to 5, where:
0: No apparent modification in the schema, even a minimal one.
1: Poor update, poorly coherent with the modification you gave me.
2: Contains significant issues, but a partial attempt of update is visible.
3: Adequately written, but lacking clarity, completeness according to the chosen modification.
4: Well-written, mostly coherent with the intended schema modification task, with only minor issues or inconsistencies.
5: Excellent, fully coherent with the schema modification task, structurally consistent, and clearly written

  {Updated schema}
\end{lstlisting}

\section{Generate Prompts}\label{app:generate}

The following prompts were submitted to \schemalink for the generation of schemas from scratch in the considered use cases:
\begin{itemize}
    \item {\tt Disease}: 
    {\it Generate a LinkML schema involving diseases.}

    \item {\tt Drug}: 
    {\it Generate a LinkML schema involving drugs and their mechanisms of action.}

    \item {\tt Protein}: 
    {\it Generate a LinkML schema involving proteins, including functional domains and post-translational modifications.}

    \item {\tt Pathway}: 
    {\it Generate a LinkML schema involving biological pathways and their participating entities (genes, RNAs, proteins, metabolites).}

    \item {\tt RNA}: 
    {\it Generate a LinkML schema involving relations between RNA molecules (including subtypes such as ncRNA, mRNA, and subtypes such as miRNA, siRNA, etc.).}
\end{itemize}

\section{Intelligent Operations and Collections}\label{app:operations+collections}

Supplementary Table~\ref{tab:collezioneAssociataOperazione} reports the retrieval collection that achieved the highest average score according to the LLM-judges for each intelligent operation. The column {\tt Avg} indicates the mean score assigned by the judges, whereas the total number of invocations is reported in {\tt Times}.
In 18 cases ($\approx$42\%), the most effective collection belongs to one of the specialized subsets $C$, $R$, or $C\!+\!R$.
Ten operations (e.g., {\it FixClassExamples}, {\it AddSubschemaExamples}) are not included in the table because they were never selected by any LLM-judge during the experimental evaluation. We associate them with $(S)$.

\begin{figure}[ht]
    \centering
    
    \includegraphics[width=0.5\linewidth]{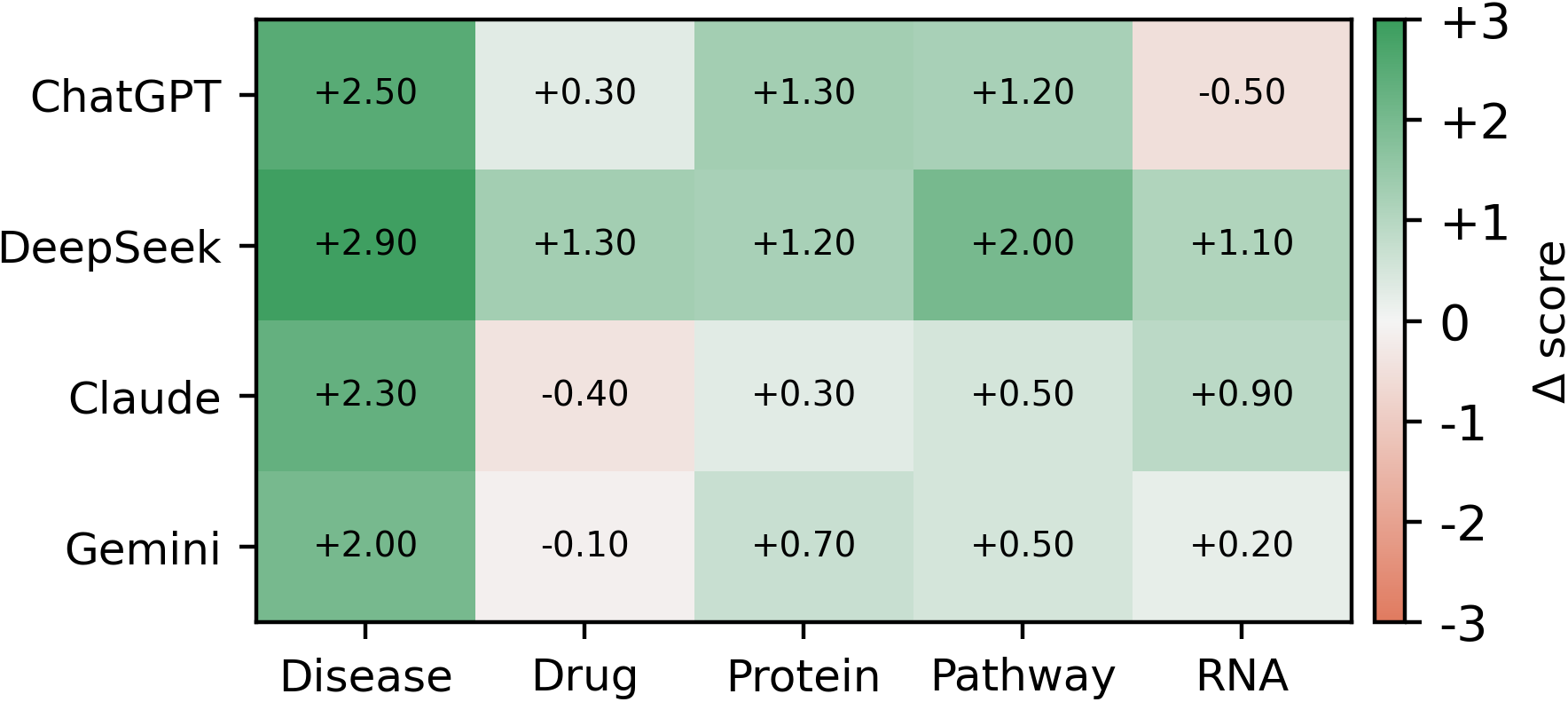}

    $(C)$

    \includegraphics[width=0.5\linewidth]{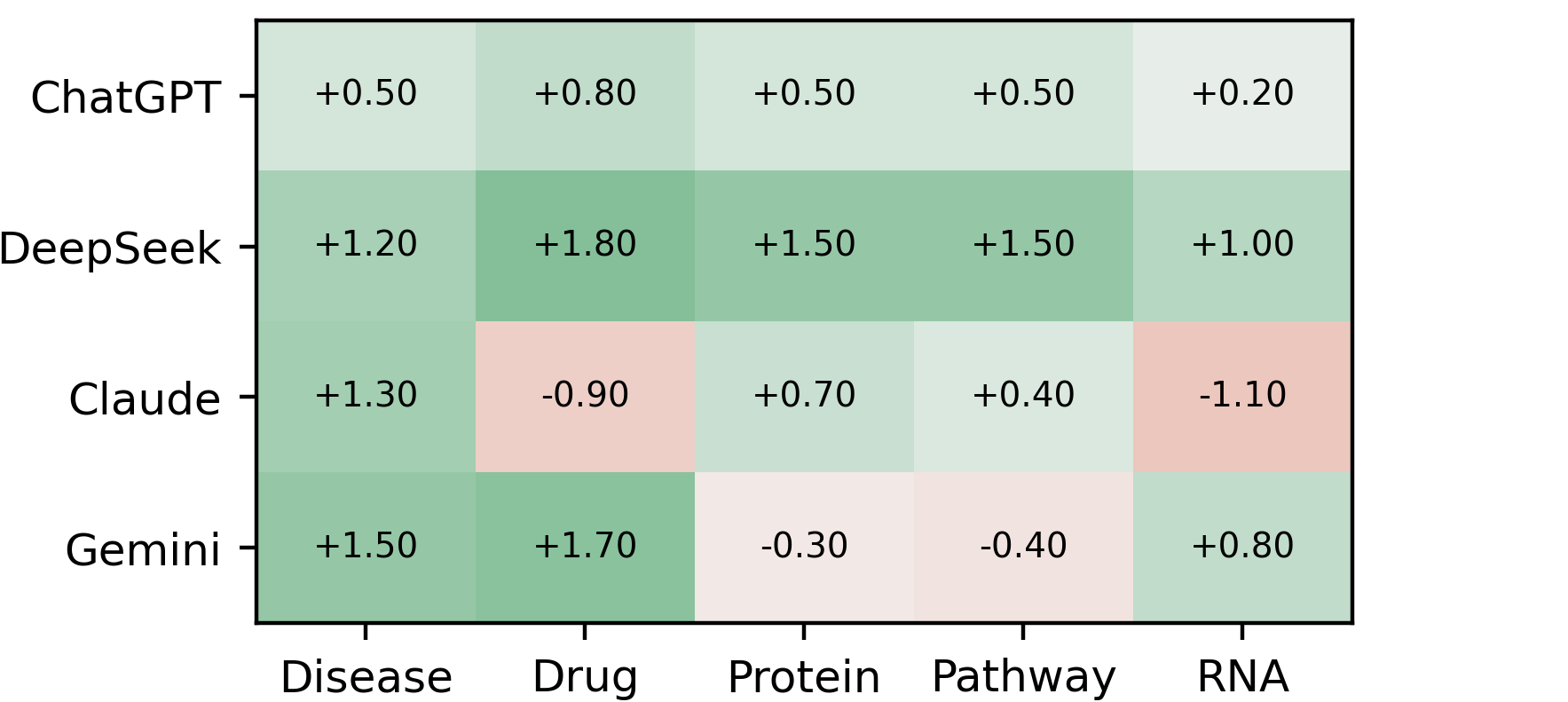}

    $(R)$

    \includegraphics[width=0.5\linewidth]{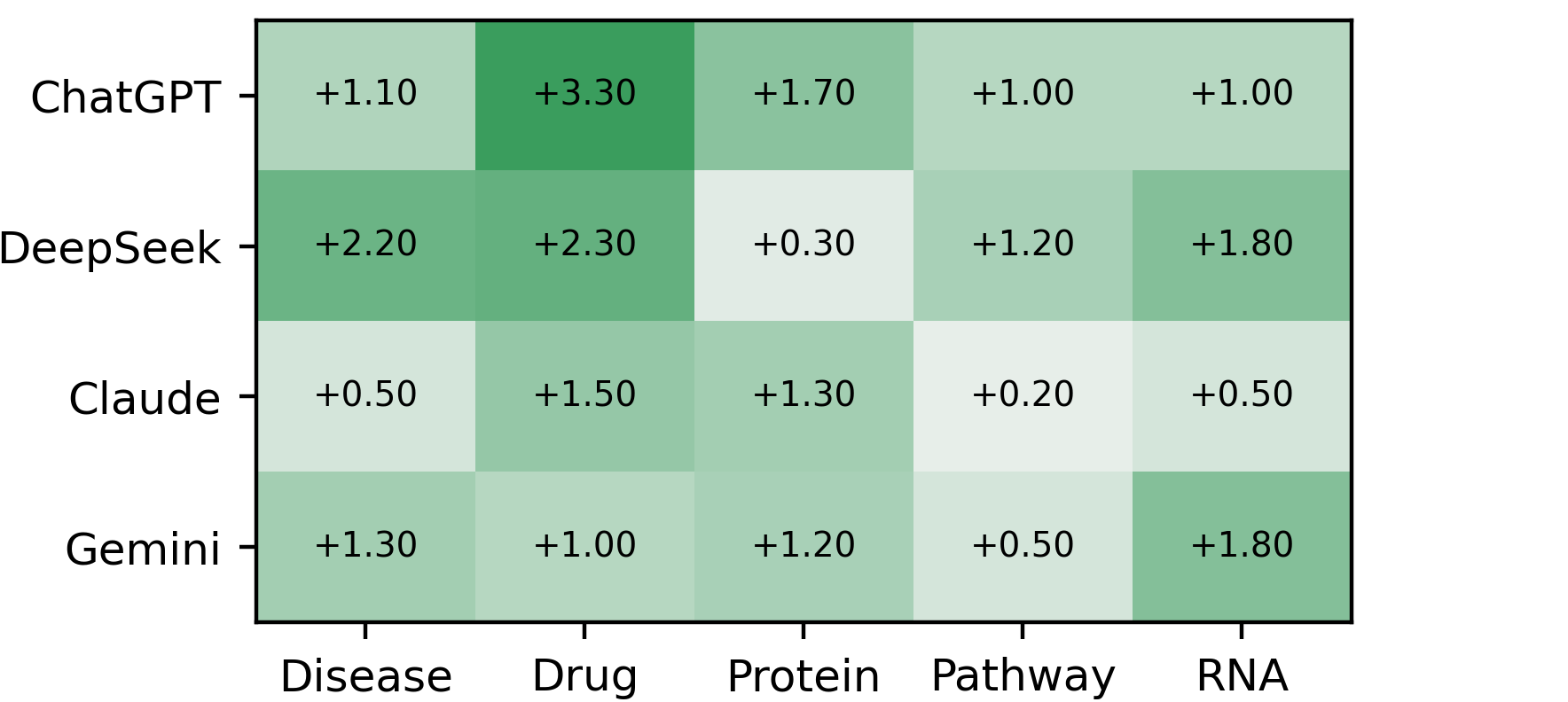}

    $(C\!+\!R)$
    
    \caption{Impact of our custom RAG collections wrt.\ $(C)$, $(R)$, and $(C\!+\!R)$.}
    \label{fig:customwrtothers}
\end{figure}

Finally, Supplementary Fig.~\ref{fig:customwrtothers} reports the average score differences obtained by using the ``custom'' collection strategy with respect to the individual retrieval policies $(C)$, $(R)$, and $(C\!+\!R)$. 
The same trend observed in the comparison against $(S)$ is confirmed: selecting the most appropriate collection for each editing operation outperforms on average single fixed retrieval strategy.

\begin{table}[ht]
\centering
\small
\begin{minipage}{0.45\textwidth}
\begin{tabular}{lccc}
\hline
\textbf{Operation} & \textbf{Coll.} & {\bf Avg} & {\bf Times} 
\\
\hline
Generate & {\tt S} & 3.5 & 20 \\\hline
AddClassSimilarToClass & {\tt S}  & 3.3 & 3 \\
AddClassAssociatedWithClass & {\tt C}  & 2.9 & 11 \\
AddAttributesToRelationship & {\tt C+R}  & 4.2 &6 \\
AddClassesSimilarToEntities & {\tt C+R}  & 3.3 &3 \\
ReifyClass & {\tt S}  & 3.0 &1 \\
FixClassName & {\tt R}  & 2.8 &8 \\
FixClassOntology & {\tt S}  & 3.0 &3 \\
FixRelationshipCardinality & {\tt S} & 2.9 & 12 \\
AddAttributesToClass & {\tt S}  & 4.0 & 8 \\
AddAttributesDescription & {\tt S}  & 4.5 & 2 \\
AddParentClass & {\tt S}  & 3.0 & 4 \\
AddChildClass & {\tt S}  & 3.7 & 6 \\
AddClassOntology & {\tt C}  & 4.6 & 7 \\
AddClassExamples & {\tt C} & 4.4 & 5 \\
AddClassDescription & {\tt C}  & 4.8 & 16 \\
FixClassDescription & {\tt R}  & 3.9 & 9 \\
FixClassAttributesName & {\tt R}  & 4.0 & 2 \\
FixClassAttributesDescription & {\tt R}  & 5.0 & 1 \\
FixClassAttributesType & {\tt C+R} & 0.63 & 8
\\
AddRelationshipAttributesDescription & {\tt S} & 5 & 1 \\
AddRelationshipOntology & {\tt C+R} & 4.3 & 4 \\
AddRelationshipExamples & {\tt C+R} & 5.0 & 2 \\
AddRelationshipDescription & {\tt C+R} & 4.8 & 5 \\
FixRelationshipName & {\tt C+R} & 2.0 & 1 \\
FixRelationshipDescription & {\tt S} & 3.0 & 1 \\
FixRelationshipAttributesName & {\tt S} & 4.0 & 1 \\
FixRelationshipAttributesType & {\tt S} & 1.0 & 3 
\\
AddAssociationsSimilarToEntities & {\tt C+R} & 4.0 & 2 \\
AddSubschemaDescription & {\tt C+R} & 2.0 & 1 \\
FixClassesAndAssociationsName & {\tt S} & 2.0 & 1 \\
FixClassesAndAssociationsDescription & {\tt R} & 4.0 & 1 \\
FixSubschemaOntology & {\tt S} & 5.0 & 1 \\
\\
\end{tabular}
\end{minipage}
\caption{Best-performing retrieval collection for each operation.}
\label{tab:collezioneAssociataOperazione}
\end{table}

